# Starving accretion disks in the center of galaxies: The examples of Sgr A* and M31

Heino Falcke[1] and Olaf M. Heinrich[2]

[1] Max-Planck-Institut für Radioastronomie, Auf dem Hügel 69, D–53121 Bonn, Germany
[2] Institut für Theoretische Astrophysik der Universität Heidelberg, Im Neuenheimer Feld 561, D-69120 Heidelberg, Germany

**Abstract.** Observations of the Galactic Center source Sgr A* suggest the presence of a central hot accretion disk around a supermassive black hole with an accretion rate a factor $10^5$ below the Eddington limit. We argue that purely spherical accretion models have problems with obervational constraints. The low accretion rate inferred for Sgr A* requires a distinct type of accretion disk – a starving disk. We investigate the structure of this type of disk within the framework of thin Keplerian disk models taking into account relativistic effects, detailed opacitiy tables and turbulent heat transport. It appears as if starving disks are the perfect realization of geometrical thin, optically thick disks with well behaved structure functions. Although in general gas-pressure dominated, radiation pressure may become important in the inner region of the disk. All in all starving disks are very stable structures with high surface density, high optical depth and long viscous time scales. However, S-curves in the $\Sigma - \dot{M}$ diagram occur, indicating a thermal limit-cycle behaviour observable in the IR-NIR with timescales ranging between three years and some thousand years. We propose that, at least in Sgr A*, the main mass supply for the disk might be a wind from the surrounding star cluster. As the disk is a very stable and slowly evolving structure, it will inhibit spherical accretion and incorporate the inflowing matter in the general accretion flow. On the other hand self-gravitation and fragmentation of the outermost part of the disk could have resulted in starformation at 0.1 pc and hence even the wind of the star cluster can already have substantial angular momentum. The situation in M31 may be very similar but with a more extreme set of parameters.

**Key words:** accretion disks – hydrodynamics – Galaxy: center – galaxies: active – galaxies: individual: M31– galaxies: nuclei

## 1. Introduction

In recent years the interest in the physical processes happening in the center of our galaxy has grown considerably, especially because of possible similarities between the Galactic Center (GC) region and the nuclei of active galaxies (AGN). It is believed that the very center of the Galaxy hosts a supermassive black hole with mass $M_\bullet \simeq 10^6 M_\odot$ (e.g. Genzel & Townes 1987). This is strongly supported not only by mass estimates but also by the presence of the compact flat spectrum radio source Sgr A* with size $\lesssim 10^{14}$ cm $\simeq 1$ mas (Krichbaum et al. 1993) at the dynamical center of the central starcluster (Eckart et al. 1993) which could be interpreted as the core of a jet-like wind originating from a central accretion disk (Falcke et al. 1993b). The presence of an accretion disk around a supermassive black hole is also consistent with NIR measurements of this region (see Falcke et al. 1993a and refererences therein).

Black hole, jet and disk could make a perfect active nucleus in the Galaxy but the actually observed signs of activity in this region are not very impressive. Generally this is attributed to a very low central accretion rate of the order $\lesssim 10^{-7} M_\odot$/yr making the GC, if at all, only an AGN on a starvation diet (Falcke & Biermann 1994a). For a black hole the critical Eddington accretion rate is a factor $\sim 10^5$ higher than this value. Usually nuclear disks around black holes are studied only in a regime close to being critical while the putative disk around Sgr A* has to be extremely subcritical.

If we believe that our galaxy is not a very special one in the universe one should think that the configuration of a starving disk plus the remnant of a radio jet can be found also in the nuclei of other ordinary spirals. Alas, the characteristic of starving disks – its low accretion rate – will make them almost invisible in more distant galaxies. At least our neighbour M31 does show in its nucleus a weak central radio source (Crane, Dickel & Cowan 1992) and indication for the presence of a central black holes (Dressler & Richstone 1988; Kormendy 1988). Therefore starving disks may be a frequent phenomenon in the center of galaxies.

In this paper we will first argue in Sec. 2 that 'disk accretion', i.e. implying the presence of an accretion disk with a large outer radius and a radio jet, is a more likely scenario for Sgr A* than recently proposed spherical accretion models where the radio emission is produced during a mostly spherical infall. Nevertheless, a reconciliation of disk and wind accretion seems possible by invoking a pre-existing (fossil) disk. Moreover, we investigate the structure of starving disks in more detail based on a code developed by Heinrich (1994) using the basic Shakura & Sunyaev (1973) $\alpha$-Ansatz with relativistic corrections for the Kerr metric (Novikov & Thorne 1973) but with refined opacities for temperatures above $\sim 2000$ K and inclusion of turbulent heat transport. The basic equations are briefly repeated in Section 3 and results for the example of Sgr A* are presented in Sec. 4. As an alternative parameter set we discuss in Sec. 5 shortly a situation applicable to M31.

## 2. Spherical accretion vs. disk accretion

### 2.1. Arguments against purely spherical accretion in Sgr A*

One has to admit that there are alternative models to explain the phenomena seen in Sgr A*. There are two main scenarios competing with each other. One is the standard paradigm for AGN scaled down to the Galactic Center with black hole, jet and accretion disk (Zylka et al. 1992; Falcke et al. 1993a&b; Falcke 1994a&b). Here it is assumed that the matter accretes through a disk structure onto the black hole, part of this matter is then expelled again by a jet formed in the very inner parts of the disk close to the black hole.

The other model – the spherical wind accretion scenario – put forward by Melia (1992a & 1994), Melia et al. (1992) and Ozernoy (1989) assumes that Sgr A* accretes in a Bondi-Hoyle type accretion from the wind of the IRS 16 complex, a small subcluster of NIR bright stars with possible temperatures of some 10,000 K. Radio and X-ray emission is then produced in onion-like shells around the central black hole with the highest frequencies produced close to the black hole. The spectral characteristics of this accretion process are then used to obtain mass estimates for Sgr A*, but despite using similar approaches both authors come to contrary conclusions. Melia, using a more elaborate description of the magnetic bremsstrahlung, confirms a supermassive balck hole of mass $\sim 10^6 M_\odot$ while Ozernoy denies it and allows only $10^3 M_\odot$ for the central mass. In the disk models a value of $10^6 M_\odot$ is favoured.

The main difference between the spherical and the disk scenario therefore does not concern the central mass but the accretion rate. Melia needs a very high accretion rate which is $10^{-4} M_\odot$/yr. He argues that the spherical infall will still have a net angular momentum which will lead to the formation of a small disk inside a certain radius. Thus he is able to account for the NIR source at the position of Sgr A*; to reduce the efficiency of the black hole in producing the visible luminosity Melia considers only a Schwarzschild hole, but still with the high accretion rate the total disk luminosity would be well above the observed limits for the luminosity of the central object. He therefore presents the spectrum of a highly inclined Newtonian disk which brings down the apparent disk luminosity by more than a factor 10.

Disregarding that relativistic effects (i.e. boosting and light bending) would diminish this effect drastically the comparison of apparent disk luminosity and observed luminosities is highly misleading. The luminosity estimates from the central source are obtained by making use of the fact that most of the nuclear luminosity at optical wavelengths gets absorbed in the dust of the central region and is reradiated in the IR, therefore one has to use *total* and not *apparent* luminosity of the disk for the comparison. Inclination effects will make no difference in this number. Falcke et al. (1993a) and Zylka et al. (1992) have shown that from the IR luminosities and taking the stellar light contributions into account, Sgr A* can not have a total luminosity $L_{\rm tot}$ larger than $\sim 7 \cdot 10^5 L_\odot$ which is also consistent with estimates for the NIR source. If one relaxes all limits and postulates clumpiness and a special geometry of the dust distribution Davidson et al. (1992) get as a very strict upper limit $L_{\rm tot} < 10^7 L_\odot$ for the nuclear luminosity which still is almost an order of magnitude below the luminosity which would be produced by central disk in the Melia model ($\sim 10^8 L_\odot$) and a rotating black hole would make this even worse. Hence, in order to reproduce the radio and X-ray emission of Sgr A* correct the spherical accretion model of Melia (1992a & 1994) is in serious conflict with observational constraints in the energetically dominant NIR-UV regime. This discrepancy is slightly reduced if the outer disk radius is very small, however, it must be extremely low ($\lesssim 5 R_{\rm s}$, $R_{\rm s} = 2GM/c^2$) such that the disk is almost non-existing to resolve this problem completely.

### 2.2. Fossil disk and wind accretion

Now, we do not deny that there are indications for a strong wind in the Galactic Center (e.g. Yussef-Zadeh & Melia 1992) but first of all the inferred accretion rate of Sgr A* depends very sensitively on the wind parameters at the postion of Sgr A*, the overall geometry of the wind and the real distance between wind source and black hole. Secondly, the question is if the wind has indeed the parameters quoted by Melia will this wind affect the accretion rate in the energetically important innermost region?

Imagine that there is a fossil disk in the center of the Galaxy – a remainder of past activity or even of galaxy formation. In how far does it influence current accretion flows? If there is no significant mass supply for the central region from larger scales those fossil disks will appear as the starving disks discussed in this paper. As we will show these disks are very stable and slowly evolving structures that may last for a long time. The disks themselves still have a large amount of angular momentum, mass and kinetic energy. A spherically infalling wind will sooner or later hit the disk which definitively will lead to a strong distortion and even disruption of the spherical infall. Especially in the inner regions the wind will be incorporated into the general disk accretion and mainly lead to a local enhancement of the accretion rate. It will take a long viscous time scale until this enhanced accretion rate reaches the black hole and becomes "visible".

Such a concept has several advantages, firstly, a fossil disk will preserve the memory of the initial orientation of the Galactic plane. This makes it more likely for an accretion disk like in Sgr A* to be somehow aligned with the galactic plane, as indicated by the radio observations (Krichbaum et al. 1993) although it is so small that the disk potential of the Galaxy is irrelevant to the dynamics of this region. Secondly, the problem of the mass supply in the central parsec is no longer important. A few stars could provide enough fuel to keep the central engine alive. First of all it supplies mass and secondly, the lower angular momentum of the more or less spherically infalling matter will counteract the tendency of the fossil accretion disk to transport its original angular momentum outwards. Wind accretion will simply lead to a shrinking of the fossil disk down to an equilibrium radius – the angular momentum of the disk is conserved, but the matter is increased – and a mass reservoir at the outer edges of the disk will develop. Hence, wind accretion can conserve angular momentum and orientation of a fossil disk for a long time and refill empty mass stores. Different epochs of star formation may then also lead to different stages of activity in the center of galaxies.

For Sgr A* we can make some simple arguments. The parameters assumed for a possible wind produced by nearby stars are a total mass loss rate $\dot M_{\rm w*} = \dot m_{-2.5}\, 3.5\cdot 10^{-3} M_\odot/{\rm yr}$, a wind velocity $V_{\rm w*} = v_{600}\, 600$ km/sec and a distance of the wind sources from Sgr A* of $D_* = d_{0.1} 0.1$ pc (Melia 1992a). The accretion radius is $R_{\rm acc} = 2GM_\bullet/V_{\rm w*}^2$ which is for a black hole of mass $M_\bullet = m_6\, 10^6 M_\odot$ approximately $R_{\rm acc} \approx 7.3\cdot 10^{16}\,{\rm cm}\; m_6/v_{600}^2$. The wind accretion rate onto the black hole is $\dot M_{\rm w\bullet} = 5\cdot 10^{-5} M_\odot/{\rm yr}$ and the mass flow per area through a disk with radius $R_{\rm acc}$ (disregarding inclination effects) would be $\dot\Sigma_{\rm w\bullet} \approx 1.8\cdot 10^{-13}\,{\rm g/sec/cm^2}\; \dot m_{-2.5} d_{0.1}^{-2}$. We can compare this with the radial accretion flow in the inner region $\dot M_{\rm disk}/4\pi R_{\rm disk}$ and see that the wind accretion becomes important at a scale $R_{\rm disk,out} \gtrsim 3\cdot 10^{15}\,{\rm cm}\; d_{0.1}\dot m_{-2.5}$. This is much larger than the inner hot disk which is smaller than $10^{14}$ cm with viscous timescales of the order $> 10^7$ years, thus larger than the expected lifetime of such a wind.

Of course the whole argument rests on the assumption that a cold, molecular fossil disk indeed extends to radii comparable to $R_{\rm disk,out}$ which is able to catch the wind before it reaches the small scales. The enclosed mass in the molecular disk can easily be of the order of a few 100 solar masses and should be seen at infrared wavelengths below $\lambda 18\mu m$ (see Zylka et al. 1993 for discussion of this topic). Another prediction of this concept would be an increased activity in some $10^7$ years which, unfortunately, is slightly longer than an astronomers lifetime.

## 3. The model

We now want to discuss the properties of starving disks like in Sgr A*. Here we concentrate only on the inner hot disk ($\gtrsim 2000$) K with a length scale $\lesssim 10^{14}$cm. We, however, implicitly postulate that a cold molecular disk extends further out which incorporates the stellar wind flow onto the center as described in the previous section and thus serves as a buffer between inner disk and environmental impact. This allows us to ignore the effects of time dependence and vertical mass inflow for the hot disk. Distortion due to the passage of stars is also not expected as this scale is much smaller than the core radius of the star cluster (Eckart et al. 1993). There is, however, one caveat that onw should bear in mind: to explain the radio emission as emission from a jet requires that a large amount of energy is transferred from the disk into the jet (Falcke et al. 1993b, Falcke et al. 1994) in the very inner part of the disk (Falcke & Biermann 1994b). This may considerably modify the inner boundary conditions of the disk and in particular the effective temperature at the disk surface. However, the problem of the backreaction of a strong outflow on the disk itself is not solved and can not be calculated in this paper.

Detailed models for the vertical structure of the disk under the assumption that the disk is geometrically thin and the fluid elements move on nearly circular Keplerian orbits have been discussed by a number of authors (e. g. Pringle 1981, Meyer and Meyer Hofmeister 1982, Cannizzo 1992). In the following we briefly list the structure equations of the disk model used in this paper (see also Heinrich 1994). The basic equations describe the mechanical equilibrium, heat generation, energy transport and the viscosity law in the disk:

$$\frac{dp_{\rm g}}{dz} = -\rho g_z + c^{-1}\rho\kappa F_{\rm rad}, \qquad (1)$$

$$\frac{dF}{dz} = \frac{9}{4}\rho\nu_{\rm turb}\omega^2, \qquad (2)$$

$$F = F_{\rm rad} + F_{\rm turb}, \qquad (3)$$

$$F_{\rm rad} = -\frac{16\sigma}{3\kappa\rho}T^3\frac{dT}{dz}. \qquad (4)$$

$$F_{\text{turb}} = -\rho \chi_{\text{turb}} T \frac{ds}{dz}, \quad (5)$$

where $g_z$ is the vertical gravitational acceleration, $\omega$ the angular velocity, $T$ the local temperature, $\rho$ the mass density, $\kappa$ the Rosseland mean opacity, $s$ the specific entropy and $F$ the total vertical energy flux which is the sum of the radiation flux $F_{\text{rad}}$ and the turbulent heat flux $F_{\text{turb}}$. For the turbulent viscosity $\nu_{\text{turb}}$ we adopt the $\alpha$-parametrization of Shakura and Sunyaev (1973) in the following form:

$$\nu_{\text{turb}} = \frac{2}{3} \frac{\alpha p_g}{\rho \omega}. \quad (6)$$

The energy transport by turbulent eddies is modelled according to the approach of Shakura et al. (1978) and Rüdiger et al. (1988, 1990). The ratio of turbulent heat conductivity $\chi_{\text{turb}}$ to turbulent viscosity is given by the dimensionless Prandtl number

$$Pr = \frac{\chi_{\text{turb}}}{\nu_{\text{turb}}}. \quad (7)$$

For the calculations in this paper we will assume a value $Pr = 1$. Rosseland mean opacities were calculated with a program of Hauschildt (1991) based on absorption cross-sections of Mathisen (1984) and Kurucz (1979). It includes electron scattering, free-free absorption and the bound-free transitions of H I, $H^-$, He I, He II, C I-III, N I-III, O I-III, Na I, Mg I-II, Al I, Si I-II, S I, K I, Fe I-II; solar abundances are assumed.

## 4. The Sgr A* disk

### 4.1. Structure of the disk

We have computed disk models for four different values of the viscosity parameter: $\log \alpha = -4, -3, -2, -1$. A black hole mass $M_\bullet = 10^6 M_\odot$, an accretion rate $\dot{M} = 10^{-7} M_\odot yr^{-1}$ and a canonical Kerr-parameter a=0.9982 (Falcke et al. 1993a) is assumed.

Figures 1&2 show two typical results of a vertical structure calculation. Fig. 1 correspond to a low value of the surface temperature ($T_s = 2000K$). In the vertical profiles the influence of hydrogen ionization can be clearly seen. In the inner layers where hydrogen is partially ionized, the eddy flux $F_{\text{turb}}$ is the dominant energy transport mechanism, while the atmospheric and sub-atmospheric layers are almost purely radiative. The density profile (dash – tripple dots) is rather flat in the interior, but has the usual exponential behaviour in the outer layers. Figure 2 represent a vertical structure for a section with surface temperature $T_s = 4.1 \cdot 10^4 K$. In this case hydrogen is fully ionized. The density profile is very smooth. The energy transport by radiation dominates although the turbulent heat flux still contributes 40 % near the midplane.

In the thin Keplerian disk approach the knowledge of the vertical disk structure allows one to compose a whole disk model from radial rings. In this way we have displayed in Figs. 3-8 the radial run of some important quantities.

It is useful to compare the results for starving disk models with those for disks with subcritical luminosity since the latter have been studied more extensively in the context of AGN. We have thus additionally calculated a disk model with the same black hole mass and $\log \alpha = -1$, but with an accretion rate $\dot{M} = 1.97 \cdot 10^{-3} M_\odot yr^{-1}$ corresponding to a luminosity $L/L_{\text{edd}} = 0.29$. For this model the condition of geometrical thinness is still fulfilled since one has $(h/r)_{\text{max}} \approx 0.1$ (Laor & Netzer 1989). The corresponding curves in Figs. 3-8 are labelled with "agn".

In Fig. 3 the variation of the disk halfthickness with radius is shown. The condition of geometrical thinness is very accurately fulfilled for starving disks, we found a maximal value $(h/r)_{\text{max}} = 3.5 \cdot 10^{-3}$ for the model with $\alpha = 10^{-4}$ while the absolute thickness of the disk is still substantial and comparable with stellar radii. The increase of $h$ with $r$ is nearly linear, there is no sign of thickening near the inner edge as in the AGN-case.

The vertical optical thickness $\tau$ and the surface density $\Sigma$ as function of $r$ are shown in Figures 4&5. For starving disks with low values of the viscosity parameter $\alpha$ the values of $\tau$ and $\Sigma$ are remarkably high. Lower values of $\alpha$ require higher values of $\Sigma$ in order to maintain the required mass supply. The optical depth can even exceed the values for an AGN-disk! The reason is the strong contribution of the bound-free opacity in starving disks due to the low surface temperature. A vertically averaged value $\bar{\chi}$ for the opacity in $cm^2 g^{-1}$ is given as $\bar{\chi} \approx \tau/\Sigma$. It follows that $\bar{\chi}$ is several hundred times larger in starving disks than in AGN-disks. Due to hydrogen ionization the curves deviate from monotonic behaviour near the outer edge of the displayed radial range which is particularly obvious for the higher values of the viscosity parameter $\log \alpha = -1, -2$. The opacity changes rapidly over several orders of magnitude. For radii larger than $\log r > 13.5...14$ the optical thickness is strongly reduced but the precise values depend on the not satisfactorily known molecular and dust opacities. Figure 6 gives the temperatures in the midplane of the disk. The values of $T_c$ are considerably high and exceed $10^6 K$ for small radii. Near the outer edge the ratio $T_c/T_s$ decreases due to the large decrease of the optical thickness.

The high temperatures in the inner disk region raise the question about the role of radiation pressure. In Fig. 7 the ratio $p_r/p_g$ is shown. For low values of the viscosity parameter ($\log \alpha = -4, -3$) radiation pressure is dominant in the innermost part of the disk, while for larger values of $\alpha$ the radiation pressure does not exceed the gas pressure. Thus despite the low accretion rate the starving disks modelled here are at the verge of becoming radiation pressure dominated in their inner region.

In Fig. 8 characteristic timescales for the disk are shown. We have estimated the viscous timescale according to $t_{\rm visc} \approx r/v_{\rm r}$, where $v_{\rm r}$ is the radial accretion velocity due to viscous loss of angular momentum and the sound propagation time according to $t_{\rm s} \approx r/c_{\rm s}$ where $c_{\rm s} = (P_{\rm tot}/\rho)^{1/2}$ is the local sound velocity. The timescales for starving disks are very long compared with an AGN-disk. The sound propagation time is of the order of years, while the timescale for viscous evolution ranges in dependence of $\alpha$ from several thousand to some $3 \cdot 10^6$ years.

The reason why despite of the geometrical thinness the timescales for a starving disk are considerably longer than for the AGN-disk model is as follows. The equation of continuity requires $\dot{M} = 2\pi r \Sigma v_{\rm r}$. The ratio of accretion rates is $\dot{M}_{\rm agn}/\dot{M}_{\rm starv} \approx 5 \cdot 10^{-5}$. However, as can be seen from Fig. 5, the ratio $\Sigma_{\rm starv}/\Sigma_{\rm agn}$ is larger than this value even for $\log \alpha = -1$. Thus one has $v_{\rm r,agn}/v_{\rm r,starv} \gg 1$, which means that in a starving disk a fluid element needs a much longer time for drifting inwards. In gas pressure dominated regions the velocity of sound scales roughly with $T_c^{1/2}$ and is thus also lower in a starving disk than in a subcritical disk. The difference is even bigger in the innermost region since for the subcritical disk model the high radiation pressure additionally increases the values of $c_s$.

We can now complete the comparison of starving disks and agn disks. Somewhat surprisingly, a starving disk with low viscosity parameter $\alpha$ has values of $\tau$ and $\Sigma$ not very far from that of an AGN-disk with $\alpha = 0.1$. The most important differences occur in the halfthickness $h$, the temperature and the viscous timescale.

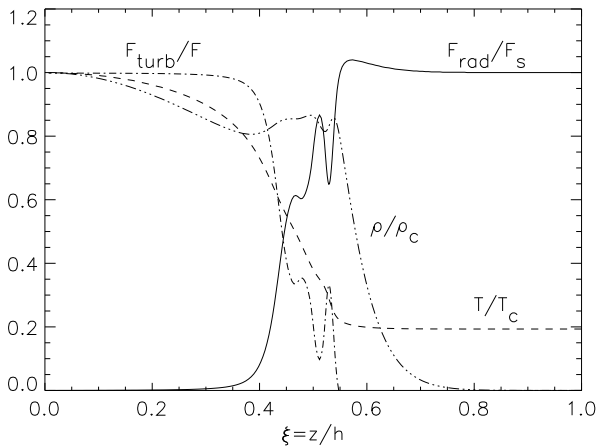

**Fig. 1.** Vertical structure for a region in the disk with low surface temperature ($T_{\rm s} = 2000K$). Density, temperature and radiation flux as well as the ratio of turbulent heat flux to total flux are plotted. Density and temperature are normalized by their midplane values $\rho_c, T_c$. The radiation flux is normalized by the flux $F_{\rm s}$ emerging from the disk surface.

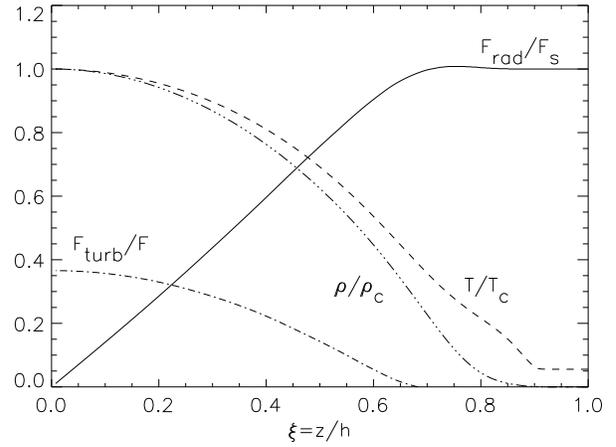

**Fig. 2.** Vertical structure for a higher surface temperature. ($T_{\rm s} = 4.1 \cdot 10^4 K$).

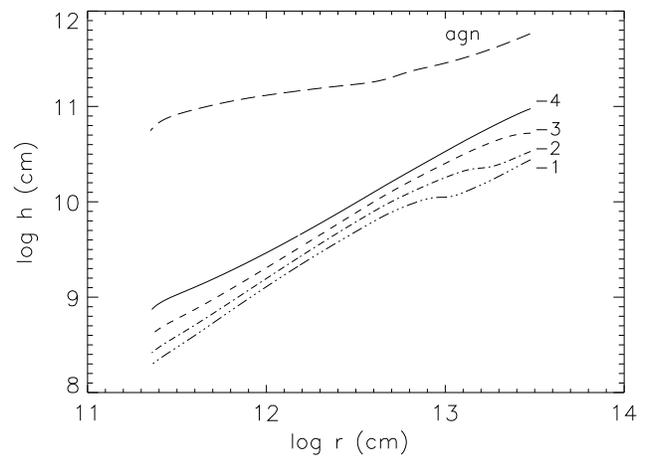

**Fig. 3.** The half thickness h of the Sgr A* disk models as a function of the radius. The curves are labelled by $\log \alpha$. Also shown is a subcritical disk model (agn) with $L/L_{edd} = 0.29$ and $\alpha = 0.1$.

### 4.2. Gravitational instability

We have seen that the disks under consideration here are relatively massive but very thin. It is thus necessary to discuss the gravitational stability of these disks. Approximately, the gravitational stability of a thin rotating disk is controlled by the dimensionless number (Toomre 1964, Lin and Pringle 1987)

$$Q = \frac{h}{R} \frac{M_{bh}}{M_d} \qquad (8)$$

where $M_{\rm d}(R)$ is the disk mass inside a radius R. The disk is gravitationally stable if $Q > 1$.

Here we have attempted to model the disk structure up to an outer radius $R_o \approx 3 \cdot 10^{13} cm$ corresponding to

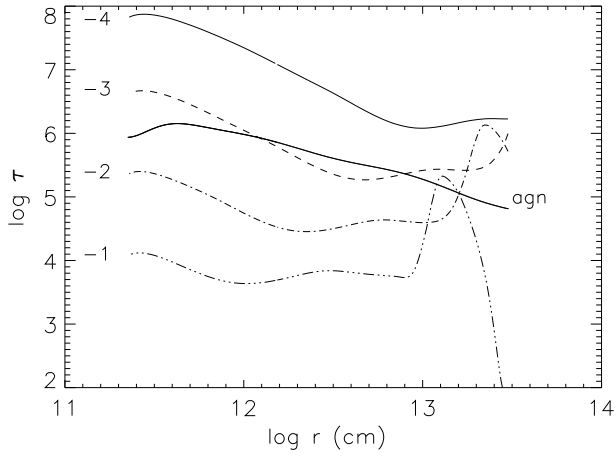

**Fig. 4.** The variation of the vertical optical depth $\tau$ with radius. (Sgr A*)

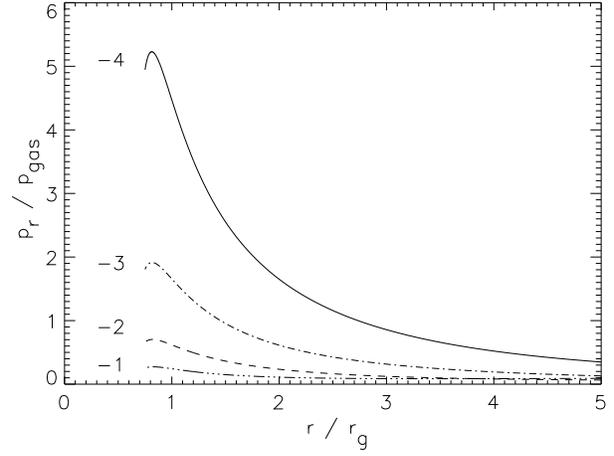

**Fig. 7.** The midplane ratio of radiation pressure $p_r$ to gas pressure $p_g$ for the innermost disk region. The radius is given here in units of the gravitational radius $r_g = 2GM/c^2$. (Sgr A*)

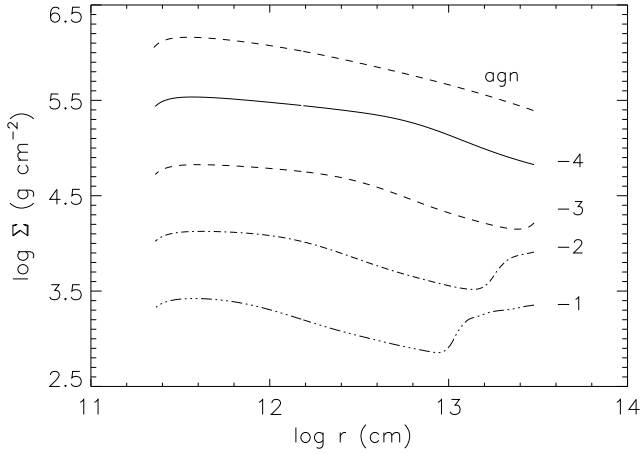

**Fig. 5.** The surface density $\Sigma$ as a function of radius. (Sgr A*)

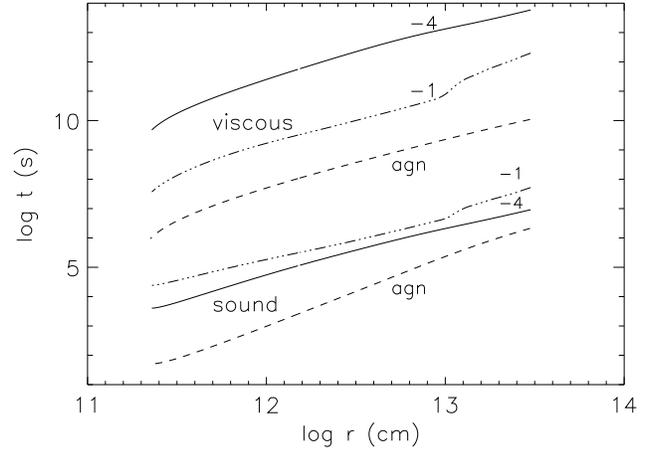

**Fig. 8.** The viscous and sound propagation timescales for the Sgr A* disk models with $\log \alpha = -1, -4$ and the subcritical disk.

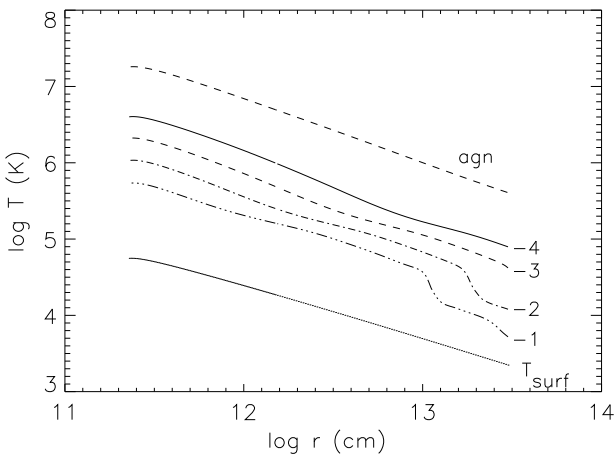

**Fig. 6.** The midplane temperature $T_c$ as a function of radius. The run of the surface temperature $T_s$ is plotted for comparison. (Sgr A*)

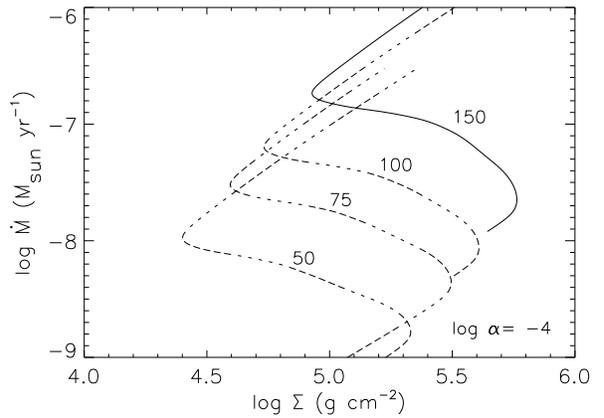

**Fig. 9.** The relation between surface density $\Sigma$ and mass flux rate $\dot{M}$ for $\log \alpha = -4$ at different radii. The curves are labelled by r in units of $r_g$. (Sgr A*)

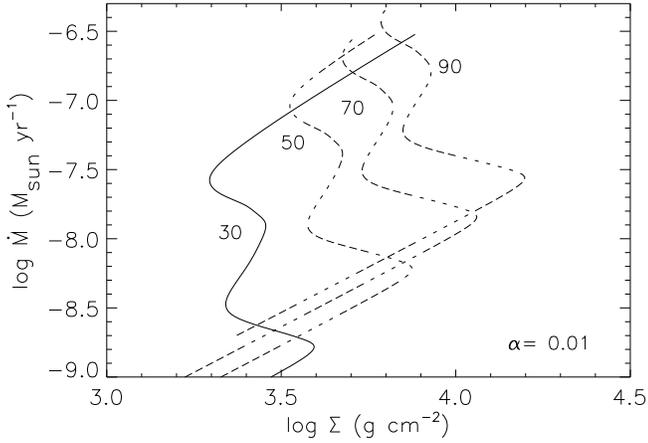

**Fig. 10.** $\Sigma - \dot{M}$ - diagrams for $\log \alpha = -2$. (Sgr A*)

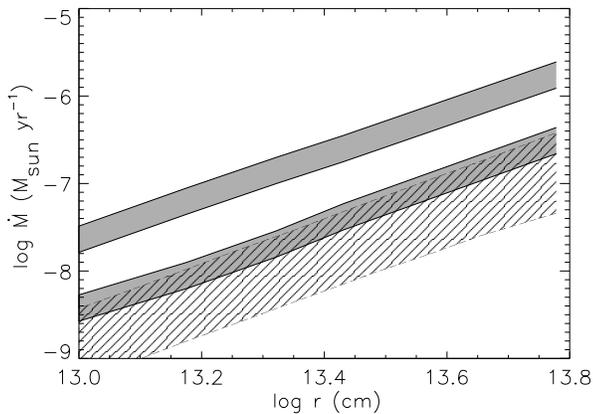

**Fig. 11.** The thermal instability strips in the $\log r - \log \dot{M}$ plane for $\log \alpha = -2$ (dark regions) and $\log \alpha = -4$ (hatched region). (Sgr A*)

$T_s \approx 2000K$. The disk mass is estimated as $M_d \approx \bar{\Sigma} R^2$, where $\bar{\Sigma}$ is an averaged value for the surface density. With $\bar{\Sigma} \approx 10^5 \mathrm{g\, cm^{-2}}$ this gives $M_d(R_o) \approx 0.05 M_\odot$ while one has $h/R > 3 \cdot 10^{-3}$. Therefore we have $Q > 6 \cdot 10^4$ and the disk region modelled here is certainly gravitationally stable.

The selfgravitation of the disk is expected to become important at a radius at least 2 orders of magnitude larger than the above value of $R_o$ hence at $R \gg 10^{16} cm$. At present it is not clear what happens in a selfgravitating disk. There are suggestions (Paczyński 1978, Kozłowski et al. 1979, Lin and Pringle 1987) that a stabilization could take place in the nonlinear regime due to a gravitationally induced viscosity, roughly given by

$$\nu_{grav} \approx Q^{-2} h c_s \qquad (9)$$

which always exceeds the $\alpha$-viscosity for $Q < 1$. In this case the outward angular momentum transport can take place at a reasonable rate in the unstable region.

Another possibility which is worth to be considered in the context of starving galactic disks has been outlined by Kolykhalov and Sunyaev (1980). One can imagine a disk where the outer parts are fragmented into stars and gas clouds forming a spinning ring around the gravitationally stable inner disk region. Material, e.g. from stellar winds, could fall onto the disk over a broad range. While in the inner region angular momentum is carried by viscous stresses and inflow of matter takes place, in the outer parts the excess angular momentum is carried mainly by a stream of matter to the fragmented torus.

Such a starforming disk picture was already suggested by Zylka et al. (1993) for the Galactic Center. And in fact one finds a small population of unusual He I emission line stars (Krabbe et al. 1991) at a distance of 0.1 pc away from Sgr A* part of which form the so called IRS 16 complex. Tamblyn & Rieke (1993) explained IRS 16 as result of a starburst some $10^7$ years ago but added a note that recent observations imply that the IMF in this area must have been very unusual — if this are stars at all. One now is tempted to link the existence of these stars to the fragmentation of the inner disk. This would have the consequence that the stellar wind produced already has a large angular momentum making purely spherical accretion even more unlikely.

### 4.3. Thermal instability

The thermal behaviour of the disk is determined by the relation between surface density $\Sigma$ and mass flux rate $\dot{M}$ at a given radial location. The concept of thermal instabilities in accretion disks which arises due to a hysteresis-like $\dot{M}(\Sigma)$-relation is now well established in accretion theory (e. g. Meyer and Meyer-Hofmeister 1981, 1983 Bath & Pringle 1982, Smak 1984). Thermal variations in low luminosity AGN-disks have been discussed by Clarke and Shields (1989).

We have calculated the $\dot{M}(\Sigma)$-relation for $\log \alpha = -2, -4$ at different radii (Figs. 9, 10). While for $\log \alpha = -4$ a simple S-shape results one has for $\log \alpha = -2$ a more complicated, double S-like relation. The branches with $\partial \dot{M}/\partial \Sigma < 0$ are thermally unstable. The occurence of double S-shapes is familiar from the study of AGN-disks (Cannizzo 1992) and caused by the fact that both a peak in the opacity mountain as well as the eddy heat transport can produce unstable branches in the $\Sigma - \dot{M}$ diagram.

Since starving disks are cooler than AGN disks the thermal instability occurs much nearer to the center. The condition $\partial \dot{M}/\partial \Sigma < 0$ defines instability strips in the $\log r - \log \dot{M}$ plane which are shown in Figure 11. For a given mass flux rate which is physically determined by the feeding mechanism at larger scales the thermally unstable radial zones can be found from Figure 11. While for $\log \alpha = -4$ there is a single instability strip, two smaller instability regions occur for $\log \alpha = -2$.

According to Meyer (1984), the thermal instability drives heating and cooling fronts through the unstable disk region. Since the front velocity is $v_f \approx \alpha c_s$, the crossing time for a front is $t_f \approx t_s/\alpha$. In the hydrogen ionization zone ($r \geq 10^{13}$cm) one has $t_s$ in the range of $10^7...10^8$s and thus $t_f$ may range from 3 to $3 \cdot 10^4$ years. As was pointed out by Clarke and Shields(1989) the spectrum of a low-luminosity disk will vary mainly in the infrared during a limit cycle. Thus, if one has $\alpha \geq 0.1$ for the Sgr A disk, an observational verification of thermal variations is in principle possible, although still very difficult in practice.

## 5. The situation in M31

In the introduction we suggested that starving disks may be common objects in the center of ordinary galaxies but that the luminosities of such disks are so low that it is really difficult to detect them. The best chance to find another sytem like in Sgr A* in another galaxy is our neighbour M31. It was discussed that in the center of this galaxy also a massive black hole of some $10^7 M_\odot$ dominates the dynamics of the inner region (see Bacon et al. 1994 and refs.). Because of the presence of a weak radio source in the dynamical center it is also argued that a similar process as in Sgr A* may be important in the nucleus of M31 as well (Melia 1992b). In the case of M31 we are not blocked by absorption in the Galactic Plane so that direct HST observations may provide reasonable constraints for the central engine.

Crane et al. 1993 and Lauer et al. 1993 presented such observations of the nucleus of M31. So far a clear pointsoure is not detected. The brightness profile extends up to $13.7^{\pm 0.1}$mag/sq.arcsec with a pixel size of $0.044''$. If we convert this into an absolute magnitude for an M31 distance of 750 kpc, we obtain $-10.7$ mag/sq.arcsec and using the emiprical relation derived by Falcke et al. (1994) this converts into a disk luminosity of $\lesssim 1.6 \cdot 10^{38} erg/sec$ when considering a point source within the central $3 \times 3$ pixel box. For the maximaly rotating Kerr hole we use here this translates into an accretion rate of $\lesssim 0.9\,10^{-8} M_\odot$/yr. This estimate should be considered as a crude upper limit but as the radio flux of the point source in M31 is a factor of 5 lower than in Sgr A* a somewhat lower accretion rate in M31 compared to Sgr A* is not unrealistic.

To highlight the principal differences in structure of the starving disks in Sgr A* and the nucleus of M31 we now simply adopt the parameters $M_\bullet = 10^7 M_\odot$ and $\dot{M} = 10^{-8} M_\odot$/yr and perform the same calculations as for Sgr A*. The variation of the vertical optical depth along the disk is shown in Fig. 12. Since the putative M31-disk is cooler by a factor of $10^{3/4}$ than the Sgr A* disk, the hydrogen-recombination region is located nearer to the inner edge when measured in units of $r_g$. The disk becomes optically thin for $r \geq 20...40 r_g$. In this transition region there is an increase of the surface density $\Sigma$ before it decreases outward again (Fig. 13). Correspondingly, the viscous timescale is prolonged to very large values (Fig. 14) and may approach a Hubble time in the outer parts. These values indicate that the turbulent viscosity is probably not efficient enough in transporting the angular momentum for radii $r \gg 10^{14}$cm. This is especially true in the regime where self-gravitation is not yet important. In order to model this part of such a low-luminosity disk one either has to abandon values of $\alpha$ as low as $10^{-4}$ or invoke other angular momentum transport mechanisms for which magnetic fields and tidal waves are good candidates. On the other hand it could be possible that the accretion ceases completely leaving a rotating ring at the outer edge.

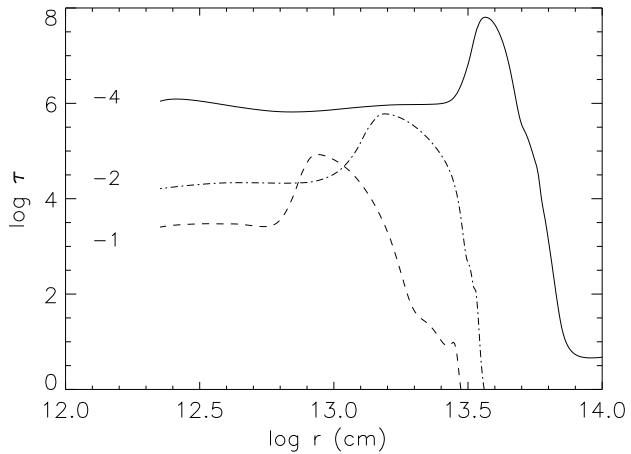

**Fig. 12.** The variation of $\tau$ with radius for the M31 disk models with $\log \alpha = -4, -2, -1$.

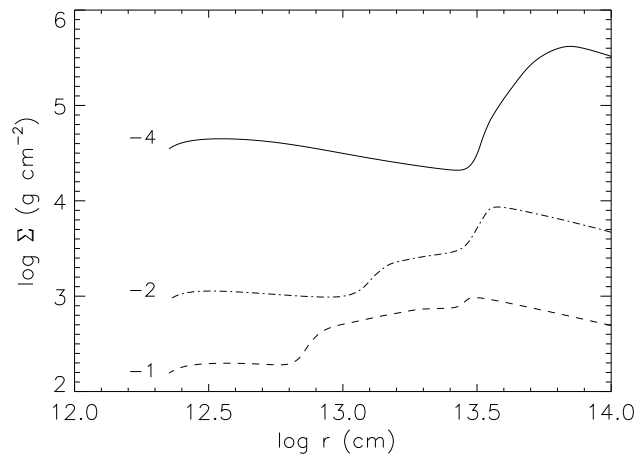

**Fig. 13.** The surface density $\Sigma$ as a function of radius for the M31 disk models.

## 6. Summary

In this paper we argue that starving disks arround black holes may be a common phenomenon in the center of galaxies. These are disks around supermassive black holes with accretion rates several orders of magnitude below the Eddington limit. We suggest that they might be fossil remains of a past and active disk accretion onto the center of the galaxy and therefore keep the memory on the original orientation of a larger disk. Even though the mass supply on larger scales down to the inner parsec might have been interrupted, we propose that the wind of stars in the central star cluster can prevent the disk from complete dissolution. The more or less spherical wind can conserve the angular momentum of the fossil disk and lead to increase of its mass reservoir. The timescale on which the accreted wind reaches the central black hole may be considerably longer than the freefall time and be dominated by the long viscous timescale of the disk. Such a scenario may explain why Sgr A* has such a low disk accretion rate (Falcke et al. 1993a) in its inner region while the inferred wind accretion rate should be much higher (Melia 1992a). Principally our approach differs from the Melia (Bond-Hoyle) model by postulating a separate inner boundary condition for the accretion flow due to a fossil disk. Without such a fossil disk the infalling wind should most likely produce a substantial central luminosity which is excluded observationally. Another possibility would be that the accretion disk has fragmented in its outer parts and formed stars. The stellar winds than would have substantial angular momentum themselves unless the isotropization timescale of the star cluster is much smaller than the (unknown) age of the present stars.

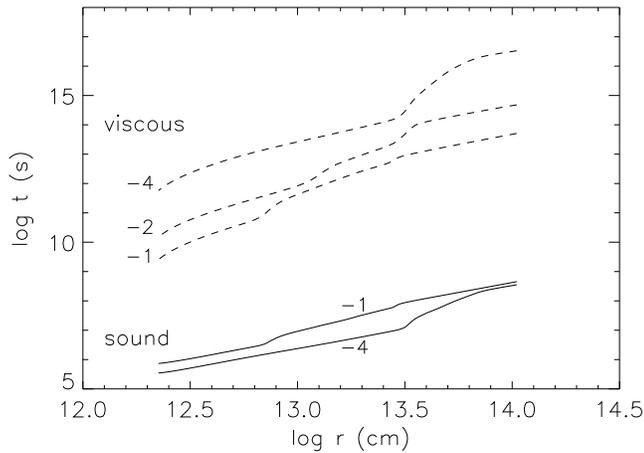

**Fig. 14.** The viscous and sound propagation timescales for the M31 disk models

We have outlined the physical properties of starving disks. Such disks are geometrically perfectly thin, but optically very thick in their inner region. Compared to AGN disks starving disks seem to be the perfect realization of the Shakura & Sunyaev (1973) accretion disk. The timescale for the viscous evolution of such disks is found to be very long (up to $10^7$ yr). The dominant source of opacity in these disks are bound-free transitions. The ra-diation pressure can be important, when the black hole mass is relatively small and the viscosity is very low. In general there are thermally unstable radial zones in which a hysteresis-like surface density-mass flux relation occurs. The corresponding limit cycles can have periods in the range from several years up to several thousand years and could be seen in the IR-NIR regime.

The same model as developed for Sgr A* may apply for the center of M31 as well. However, because of an even lower accretion rate onto a probably more massive black hole the situation is more extreme. The 'hot' disk will be very small (some 20 $r_g$) and the viscous time scales may become even longer than in Sgr A*.

*Acknowledgements.* We thank F. Meyer who outlined the concept of a fossil disk and mentioned its relevance for the wind accretion scenario to us. The referee, M. Rees, pointed out the possibility of self-gravitation and star-formation. P.G. Mezger, P.L. Biermann and W.J. Duschl stimulated our interest in this field and contributed in various ways to this and the preceding work. Interesting discussions with F. Melia helped to clarify several points. The work was supported by the DFG (Bi 191/9 & SFB 328).